%Paper: hep-th/9210069
%From: mickelss@sun1.ruf.uni-freiburg.de (Jouko Mickelsson)
%Date: Tue, 13 Oct 92 14:24:31 MET

\input amstex.tex
\documentstyle{amsppt}
\magnification\magstep1

\topmatter \title HILBERT SPACE COCYCLES AS REPRESENTATIONS\\ OF (3+1)-D
CURRENT ALGEBRAS\footnote{Supported by the Alexander von Humboldt foundation}
\endtitle
\author Jouko Mickelsson\endauthor
\affil Department for Theoretical Physics, University of Freiburg. \\ E-mail
mickelss\@ sun1.ruf.uni-freiburg.de \hphantom{A} Permanent address: \\
Department of Mathematics, University of Jyv\"askyl\"a,
SF-40100, Jyv\"askyl\"a, Finland.\endaffil
\date{October 11, 1992}

\endtopmatter
\document
ABSTRACT  It is proposed that instead of normal representations one should look
at cocycles of group extensions valued in certain groups of unitary operators
acting in a Hilbert space (e.g the Fock space of chiral fermions), when
dealing with groups associated to current algebras in gauge theories in
$3+1$ space-time dimensions. The appropriate cocycle is evaluated in the
case of the group of smooth maps from the physical three-space to a compact
Lie group.

The cocyclic representation of a component $X$ of the current is obtained
through two regularizations, 1) a
conjugation by a background potential dependent unitary
operator $h_A,$ 2) by a subtraction $-h_A^{-1}\Cal L_X h_A,$ where $\Cal L_X$
is a derivative along a gauge orbit. It is only the total operator $h_A^{-1}
Xh_A-h_A^{-1}\Cal L_X h_A$ which is quantizable in the Fock space using the
usual normal ordering subtraction.

\redefine\Bbb{\bold}

\define\a{\alpha}
\redefine\e{\epsilon}
\redefine\b{\beta}

\redefine\o{\omega}
\redefine\O{\Omega}
\redefine\l{\lambda}

\define\RM{\Bbb R}
\define\CM{\Bbb C}

\define\gm{\bold g}

\define\<#1,#2>{\langle #1,#2\rangle}
\define\TR{\text{tr}}
\define\dep(#1,#2){\text{det}_{#1}#2}
\vskip 0.5in
1. INTRODUCTION

\vskip 0.3in
In gauge theory in $d+1$ space-time dimensions it is natural to look at the
representations of the group of smooth maps $\Cal G=Map(M,G),$ where $M$ is the
physical $d$ dimensional space and $G$ is the gauge group. Actually, in
order to construct representations such that the energy is bounded below
one has to construct a certain extension $\hat{\Cal G}$ of $\Cal G.$
When $M=S^1$ this extension is an affine Kac-Moody group and we have a
beautiful and well-understood theory of highest weight representations at
hand, [K], [PS].

When $d>1$ the relevant extensions are also known, first found in [M1], [FS]
(for the Lie algebra case) and in [M2] (for the corresponding groups); see
also [M3].  However, no \it unitary \rm separable faithful have been
constructed and in fact there is a no-go theorem of Pickrell, [Pi], which tells
us that the appropriate extension of a slightly bigger general linear group
$GL_p$ ($p$ is an integer depending on $d$, the index of a Schatten ideal)
does not have unitary separable faithful representations. This is a
discouraging
result since in the case $d=1$ highest weight representations can be
constructed
by restricting to the current group, [PS].

A representation of a group $G$ can be viewed as a special kind of 1-cocycle,
with values in a unitary group $U(H)$ of a Hilbert space $H.$ If we have in
addition an action of $G$ on a manifold $\Cal A,$ a 1-cocycle for this action
with values in $U(H)$ is a map $\o:G\times\Cal A\to U(H)$ such that
$$\o(g;g'\cdot A) \o(g';A)=\o(gg';A).\tag1.1$$
If $\o$ does not depend on $A$ we have an ordinary representation of $G$ in
$H.$
The main point of this paper is that it is physically quite natural to study
the
Hilbert space cocycles $\o$ when $G$ is the current group in $3$ dimensions.

The point of view developed in the present paper is closely related to the
Fock bundle approach in [M4]. In fact, the Hilbert cocycle results from a
special kind of trivialization of the Fock bundle over the space of vector
potentials. There is still a third approach to generalized representations of
current algebras proposed by Langmann, [L]. Following Ruijsenaars, [R], he
defines the elements of the current algebra as sesquilinear forms in an
appropriate dense domain of the Fock space. He shows that there is a
regularized
product of  forms which produces the commutator anomalies found in [MR].
On the other hand, the forms can be computed from the Fock bundle maps,
as discussed in [M5]. Thus the approach to "representation" theory developed
here for the group $Map(M,G)$ is equivalent to the two earlier methods.
Nevertheless, I hope that the reader will find the results useful 1) for
understanding the nature of the generalized representation theory of the
group of gauge transformations, 2) for a generalization to other
infinite-dimensional groups like the group of diffeomorphism on a compact
manifold and for constructing more general representations of the gauge
group.

I want to thank S. Rajeev on many useful discussions on current algebras,
Hartmann R\"omer for hospitality in Freiburg, where this research was done,
and Alexander von Humboldt foundation for financial support.

\vskip 0.5in
2. SOME BACKGROUND NOTIONS

\vskip 0.3in
Let $M$ be a compact oriented three-manifold with a fixed spin structure, $G$ a
compact Lie group, and $\Cal A$ the space of smooth 1-forms on $M$ with values
in the Lie algebra $\gm$ of $G.$ The topology and differentiable structure of
$\Cal A$ can be defined using a family of seminorms given by $L^2$ norms of
partial derivatives of the  vector potentials, but the choice of the smooth
structure does not play any essential role in this paper.

Fix a trivial complex vector bundle $M\times
V$ over $M$ with a unitary representation $\rho$ of $G$ in the fiber $V.$
A Dirac operator on $M$ consists of two parts $D^+:S^+\to S^-,$ $D^-:S^-\to
S^+$ mapping the positive and negative chirality spinor fields to each other.
We
shall fix an isomorphism between the spaces $S^+$ and $S^-,$ so we can view
$D^+$ as a map of $S^+$ to itself and similarly for $D^-.$  For any vector
potential $A\in\Cal A$ we have analogously (by "minimal coupling") a map
$D_A^+:S^+\otimes V\to S^+\otimes V.$ We shall drop the upper index "+" and
consider for definiteness only the positive chirality spinors.

Let $H=L^2(S\otimes V)$ be the Hilbert space consisting of square integrable
spinor fields (of positive chirality). If $\l$ is any real number not in the
spectrum of $D_A$ we denote by $H_+(A,\l)$ the closed subspace of $H$ spanned
by eigenvectors of $D_A$ belonging to eigenvalues greater than $\l$ and by
$H_-(A,\l)$ the orthogonal complement of $H_+(A,\l).$ As a reference plane
we fix $H_+=H_+(0,\l_0)$ for some $\l_0.$

Because of the flow of eigenvalues of $D_A$ as a function of $A$ it is not
possible to fix the vacuum level $\l$ once and for all without introducing
discontinuities in the function $A\mapsto H_+(A,\l).$

Denoting by $<\cdot,\cdot>$ the $L^2$ inner product in $H$ (linear in the
first, antilinear in the second argument) the canonical anticommutation
relations (CAR) are
$$a^*(u)a(v)+a(v)a^*(u)=<u,v>, \text{ with $u,v\in H$ }\tag2.1$$
all other anticommutators for the generators $a^*(u),a(v)$ are zero. If $W
\subset H$ is a closed subspace then there is a unique representation of the
CAR in a Hilbert space $\Cal F_W$ with a vacuum vector $\psi_W$ such that
$$a^*(u)\psi_W=a(v)\psi_W \text{ for $u\in W^{\perp},$ and $v\in W.$}\tag2.2$$
One would be tempted to set $W=W(A)=H_+(A,\l)$ as a function of the vector
potential. We would then have for each potential $A$ a Fock space $\Cal F_W$
with a vacuum (vector of lowest energy, the energy of the vacuum normalized
to zero by a normal ordering principle). However, this does not quite work
because of the discontinuity of $W(A).$

In spite of the fact that we cannot fix the vacuum in a continuous way
the equivalence class of the CAR representation is well-defined for each
$A\in\Cal A.$ For a given potential $A$ define the Grassmannian $Gr_1(A)$
consisting of all closed subspaces $W\in H$ such that the orthogonal projection
$W\to H_-(A,\l)$ is Hilbert-Schmidt. This condition does not depend on the
choice of $\l$ since there are only a finite number of linearly independent
eigenvectors of $D_A$ in any finite interval of the spectrum; thus the
difference of the projections for two different $\l$'s is of finite rank and
consequently Hilbert-Schmidt. It is known that a pair of vacuum representations
of CAR, corresponding to vacua $\psi_W,\psi_{W'}$ such that the projection of
$W$ to the complement of $W'$ is Hilbert-Schmidt, are equivalent,[A]. It
follows
that we may use any of the planes $W\in Gr_1(A)$ to define the CAR
representation so long as we are only interested in the equivalence class of
the representation.

The representations of CAR corresponding to vacua $\psi_W,\psi_{W'}$ with
$W=H_+(A,\l), W'=H_+(A,\l')$ are equivalent but the equivalence is defined
naturally only up to a phase. It is precisely this property of chiral fermions
which is responsible to the anomalies of the gauge group action, [S], [NA].
In the case of four component fermions the phase ambiguities from the two
different chiral sectors cancel and there is no anomaly.

\vskip 0.5in
3. THE HILBERT SPACE COCYCLES

\vskip 0.3in
Let $\e$ be the operator on $H$ which is $+1$ in the subspace $H_+$ and $-1$ in
the subspace $H_-.$ We denote by $GL_p$ the Banach-Lie group  of bounded
invertible operators $g:H\to
H$ with the property $[\e, g]\in L_{2p},$ where $L_{2p}$ is the Schatten ideal
consisting of operators $T$ such that $(T^*T)^p$ is a trace-class operator. The
unitary subgroup of $GL_p$ is denoted by $U_p.$  An element $g$ of $GL_p$ can
be
written in the block form
$$g=\left(\matrix a&b\\c&d\endmatrix\right)\tag3.1$$
with respect to the splitting $H=H_+\oplus H_-;$ so $a:H_+\to H_+,$ $b:H_-\to
H_+,$ and so on. The operators $b,c$ belong to $L_{2p}$ and from this follows
that $a,d$ are Fredholm operators of opposite index.

The action of a gauge
transformation $g\in \Cal G$ defined by pointwise multiplication on
square-integrable
spinor fields defines a homomorphism $T:\Cal G\to GL_p$, where $p=\frac12(1+dim
M),$ see e.g. [MR]. Since we are here concerned with the case $dim M=3$ we have
$p=2.$

We also need the $L_{2p}$ Grassmannians $Gr_p=GL_p/B_p,$ where $B_p$ consists
of operators with the lower left block $c=0.$ We can also write $Gr_p=U_p/K_p,$
where $K_p$ consists of the unitary operators with the off-diagonal blocks
$b,c$
equal to zero. It is easy to see that $GL_p$ is the maximal subgroup of $GL(H)$
which acts on $Gr_p.$

The Grassmannians $Gr_1(A)$ form a smooth fiber bundle over $\Cal A.$ Since the
base is flat this bundle is trivial. For any $A\in\Cal A$ let $P_A$ be the set
of unitary operators $g$ in $H$ such that $g\cdot H_+\in Gr_1(A).$ If $g\in
P_A$ then $gh\in P_A$ if and only if $h\in U_1.$ The fibers $P_A$ fit together
to form a principal $U_1$ bundle $P$ over $\Cal A.$ The bundle of Grassmannians
can be viewed as an associated fiber bundle, defined by the usual $U_1$ action
on $Gr_1.$

Choose a section $A\mapsto h_A$ of the bundle $P.$ We define the $U_1$ valued
cocycle
$$\o(g;A)=h_{g\cdot A}^{-1} T(g) h_A\tag3.2$$
It is easy to check that the cocycle condition (1.1) is indeed satisfied.
Let $\Cal G_o$ be the group if \it based \rm gauged transformations, that is,
elements $g\in\Cal G$ such that in a given base point $x\in M$ we have
$g(x)=1.$
The quotient $\Cal A/\Cal G_o$ is a smooth manifold and we may form the bundle
$Q=P/\Cal G_o$ over $\Cal A/\Cal G_o,$ defined by the right action in the
fibers
and the gauge action in the base. There is an alternative way to define this
bundle using the cocycle $\o.$ Define a right $\Cal G_o$ action in
$\Cal A\times U_1$ by
$$(A,g)\cdot h=\left(h^{-1}Ah+h d(h^{-1}),g\o(h;A)\right)\tag3.3$$
Then $Q=(\Cal A\times U_1)/\Cal G_o.$

The formula (3.2) should be considered as a regularization of the
nonquantizable
operator $T(g).$ Taking a derivative at the unity along a one-parameter
subgroup of $\Cal G$ one obtains the Lie algebra cocycle
$$(d\o)(X;A) = h_A^{-1} dT(X) h_A-h_A^{-1}\Cal L_X h_A,$$
where $\Cal L_X$ denotes the Lie derivative along the vector field generated by
an element $X$ in the Lie algebra of $\Cal G$ and $dT$ is the Lie algebra
representation corresponding to the group representation $T.$ Thus the
regularization on the
Lie algebra level (i.e., for charge densities) consists of two parts: 1) A
conjugation by $h_A$, 2) a subtraction by $h_A^{-1}\Cal L_X h_A.$ These two
operations do not make sense separately in the fermionic Fock space, but
combined they produce an element in the Lie algebra $\bold u_1,$ which is
quantizable.

In general, we can choose any $U_1$ valued cocycle $\o$ to define a $U_1$
bundle $Q_{\o}$ over $\Cal A/\Cal G_o.$ Two cocycles $\o,\o'$ are said to be
cohomologous if there exists a map $f:\Cal A\to U_1$ such that
$$\o'(g;A)=f(g\cdot A)^{-1}\o(g;A)f(A).\tag3.4$$
The reader should verify the following simple Lemma:
\proclaim{Lemma 3.5} The bundles $Q_{\o}$ and $Q_{\o'}$ are equivalent if and
only if the cocycles $\o$ and $\o'$ are cohomologous.\endproclaim
If $h,h'$ is a pair of trivializations of $P$ then $h'_A=h_Af(A)$ for some
map $f:\Cal A\to U_1$ and so the cocycles $\o,\o'$ defined by (3.2) are
cohomologous. Thus, as expected, the bundles $Q_{\o},Q_{\o'}$ constructed using
two different trivializations of $P$ are equivalent.

Whereas $P$ is always trivial, $Q$ will typically be nontrivial. Namely, if
$Q$ is trivial then the cocycle $\o$ has to be contractible as a map from
$\Cal G_o\times \Cal A\to U_1$ because in the trivial case $\o(g;A)=f(g\cdot A)
^{-1} f(A)$ for some $f$ and $\Cal A$ is a contractible space. Now the function
$h$ in the definition of $\o$ is contractible, again by contractibility of
$\Cal
A.$ But the embedding $T:\Cal G_o\to U_2$ is not contractible, [PS], in the
case
of a nonabelian compact gauge group $G.$

The idea for constructing the bundle $\Cal F$ of Fock spaces over $\Cal A,$
together with an (anomalous) gauge group action is to consider the Fock bundle
as an associated bundle to $P$ via a \it projective \rm representation of
$U_1$ in a standard Fock space $\Cal F_0.$ This would lead to a bundle of
projective Fock space (as explained in [S]). In order to create a true
vector bundle with the right physical properties we shall first lift the
cocycle $\o$ to a cocycle of an appropriate extension $\hat{\Cal G}$
of $\Cal G,$ with values in the \it central extension \rm $\hat U_1$ of $U_1.$

Recall from [PS] the construction of the
central extension $\widehat{GL}_1.$ This group consists of pairs $(g,q)\in
GL_1\times GL(H_+)$ such that $a(g)q^{-1}-1\in L_1,$ modulo the equivalence
relation $(g,q)\equiv (g,qt)$ for any $t\in GL(H_+)$ with $t-1\in L_1.$ The
operators $t$ are those which have a determinant. The multiplication is
defined simply by $(g,q)(g',q')=(gg',qq')$ and the unitary subgroup $\hat U_1$
is obtained by taking the intersection of $\widehat{GL}_1$ with $U(H)\times
U(H_+).$ A section near the unit element is
given by the map $g\mapsto q=a(g).$ The local 2-cocycle on $GL_1$ corresponding
to this choice is
$$\O_1(g,g')= \text{det}(a(g)a(g')a(gg')^{-1}).\tag3.6$$
Locally the group $\widehat{GL}_1$ is the product $GL_1\times \CM^{\times}$
with the multiplication
$(g,\a)(g',\a')=(gg',\a\a' \O_1(g,g')).$  Globally the group is a twisted
line bundle (the zero section removed) and cannot be defined using a single
cocycle.

The Lie algebra of $\widehat{GL}_1$ is $\bold{gl}_1\oplus\CM,$ where the
Lie algebra $\bold{gl}_1$ of $GL_1$ consists of all bounded operators with
Hilbert-Schmidt off-diagonal blocks. The commutator is the Lie algebra sum
commutator plus the  Lundberg's cocycle, [Lu],
$$c_1(X,Y)=\frac14\TR\epsilon[\epsilon,X][\epsilon,Y], \,\,\,X,Y\in
\bold{gl}_1 \tag3.7$$

For a given $g\in\Cal G$ we may choose (since $\Cal A$ is flat) a function
$$A\mapsto \hat{\o}(g;A)\in \hat U_1, \hphantom{AA}\pi(\hat{\o}(g;A)=\o(g;A),
\tag3.8$$
where $\pi:\hat U_1\to U_1$ is the canonical projection. Any two choices
differ by a $S^1$ valued function on $\Cal A.$ But the circle bundle $\hat
U_1\to U_1$ is nontrivial, [PS], and therefore there is no continuous lift
$\hat{\o}$ to the whole space $\Cal A\times \Cal G.$ If we make a choice of
the lift in some open contractible neighborhood of unity in $\Cal G$ then
$$\hat{\o}(gg';A)=\O(g,g';A) \hat{\o}(g;g'\cdot A) \hat{\o}(g';A)\tag3.9$$
for some local function $\O.$ This leads to an extension $\hat{\Cal G}$ of
$\Cal G$ by the abelian group of maps $Map(\Cal A,S^1).$ Elements of the
extension are pairs $(g,\theta),$ where $\theta:\Cal A\to \hat U_1$ is any lift
of $A\mapsto \o(g;A).$ The group multiplication
is defined by
$$ (g,\theta)(g',\theta')=(gg',\theta''), \text{ with $\theta''(A)=
\theta(g'\cdot A)\theta'(A)$}.\tag3.10$$

The extension \define\GG{\hat{\Cal G}} $\GG$ is completely determined by the
bundle $Q$, or in the other words, by the cocycle $\o.$

\proclaim{Lemma 3.11} The extensions $\GG, \GG'$ defined by the cohomologous
cocycles $\o, \o'$ are isomorphic.\endproclaim
\demo{Proof} Suppose $\o'(g;A)=f(g\cdot A)^{-1}\o(g;A)f(A)$ for some function
$f.$ Choose a lift $\hat f:\Cal A\to\hat U_1.$ We can now define a map
$$\phi:\GG\to \GG',\hphantom{Aa}\phi(g,\theta)=(g,\theta') \text{ with
$\theta'(A)=\hat f(g\cdot A)^{-1}\theta(A)\hat f(A)$ }\tag3.12$$
By the choice of $\hat f$ we know that $\theta'$ is a lift of $A\mapsto
\o'(g;A).$ Likewise, the homomorphism property is an immediate consequence of
the definitions. The inverse map of $\phi$ is obtained by replacing $\hat f(A)$
by $\hat f(A)^{-1}.\,\, \square$ \enddemo

The bundle $\Cal F$ of Fock spaces over $\Cal A$ is now defined as follows.
The free Fock space $\Cal F_0$ carries a faithful representation of $\hat U_1,$
[PS]. Set $\Cal F=\Cal A\times \Cal F_0,$ with the following action of $\GG.$
Let $(g,\theta)\in \GG.$ Then
$$(g,\theta)(A,\psi)=(g\cdot A, \theta(A)\cdot \psi)\tag3.13$$
To complete the picture we have to give the action of the Dirac Hamiltonian
$D_A$ in each fiber. By the trivialization $h_A$ the operator $D_A$ in the
one-particle space over $A$ is conjugated to $D'_A=h_A^{-1}D_A h_A$ in
the one-particle space over $0\in\Cal A.$ By the choice of $h,$ $D'_A$ is an
unbounded selfadjoint operator with the additional property that the
off-diagonal
blocks of its sign operator with respect to the polarization $H=H_+\oplus H_-$
are Hilbert-Schmidt.

After fixing a vacuum level $\l\notin Spec(D_A)$ there is a unique vacuum ray
$\CM\psi(A,\l)$ in the Fock space $\Cal F_0$ for the conjugated operator
$D'_A.$
As usual, the quantum Hamiltonian is then
$$\hat{D}'_A=\sum_n \l_n :a^*(u_n) a(u_n): \tag3.14$$
where $\{u_n\}$ is a complete set of eigenvectors for $D'_A$ with eigenvalues
$\l_n.$ The normal ordering is defined such that $:a^*(u_n)a(u_n):=-a(u_n)
a^*(u_n)$ when $\l_n <\l$ and no change of order otherwise; this means that
the energy of the vacuum is normalized to zero.

Note that the vacuum $\psi(A,\l)$ is a continuous function of the potential
$A$ only locally, because of the crossing of the eigenvalues of the vacuum
level. On the other hand, if we choose $\l=0$ we obtain a (weakly) continuous
family of quantum Hamiltonians $\hat{D}'_A.$

Whereas the principal bundle $P$ over $\Cal A$ can be pushed forward to a
bundle $Q_{\o}$ over $\Cal A/\Cal G_o,$ the same is not true for the Fock
bundle; the obstruction is precisely the nontriviality of the extension
$\GG.$

\vskip 0.5in
4. SOME COMPUTATIONS

\vskip 0.3in
In this section we want to present explicit formulas for the trivialization
$h_A,$ the cocycle $\o(g;A),$ and the abelian extension $\GG.$

To any (unbounded) selfadjoint operator $D$ in $H$ one can associate the
bounded operator $R(D)=D/(|D|+\exp(-|D|).$ The order in the product is
unimportant since a selfadjoint operator commutes with its absolute value.
$R(D)$ is "almost" the sign operator associated to $D.$ The addition of the
term $\exp(-|D|)$ in the denominator guarantees that the function $A\mapsto
R(D_A)$ is well-defined
and continuous also for operators having the eigenvalue zero.

Suppose that zero is not an eigenvalue of $D.$ Then the difference
$$\frac{D}{|D|}-R(D)=\frac{D\exp(-|D|)}{|D|^2+|D|\exp(-|D|)}\tag4.1$$
is a trace-class operator when $D$ is a Dirac operator; in particular, the
difference is Hilbert-Schmidt. We write the sign operator $F=F_A=D_A/|D_A|$
(and
similarly any other operator in $H$) in the block form
$$F=\left(\matrix F_{11}&F_{12}\\F_{21}&F_{22}\endmatrix\right)$$
with respect to the splitting $H=H_+\oplus H_-;$ for example, $F_{21}:H_+\to
H_-.$ The hermitean operator $F$ corresponds to a point on the Grassmannian
$Gr_2.$ Denoting $x=R(D_A)_{21}$ we put
$$h_A=\left(\matrix \frac{1}{\sqrt{1+\frac14 x^*x}}& \frac{-x^*/2}{\sqrt{
1+\frac14 xx^*}}
\\ \frac{x/2}{\sqrt{1+\frac14 x^*x}}&\frac{1}{\sqrt{1+\frac14 xx^*}}
\endmatrix\right).\tag4.2$$
This is unitary for any bounded operator $x.$

\proclaim{Lemma 4.3}  The difference $F_A-h_A\epsilon h_A^{-1}$ is
a Hilbert-Schmidt
operator.\endproclaim
\demo{Proof} We already know that $F_{21}-x$ is Hilbert-Schmidt. Now
$$h\epsilon h^{-1}=\left(\matrix 1-\frac{x^*x}{4}&x^*\\x&\frac{xx^*}{4}-1
\endmatrix\right)\left(\matrix 1+\frac{x^*x}{4}&0\\0&1+\frac{xx^*}{4}\endmatrix
\right)^{-1}.$$
The off-diagonal blocks of the sign operator $F$ are in $L_4,$ [M3, Chapter
12].
It follows that
$x^*x$ is Hilbert-Schmidt and therefore the off-diagonal blocks in the product
above differ
from $x$ (resp. $x^*$) by a Hilbert-Schmidt operator; thus the off-diagonal
blocks are equal to $F_{12}$ (resp. $F_{21}$) modulo HS operators. A similar
statement for the diagonal blocks follows from the property $F^2=1.$\enddemo

For any $A\in\Cal A$ and $\l$ not in the spectrum of $D_A$ the difference
between the sign operator of $D_A-\l$ and the operator $R(D_A)$ is
Hilbert-Schmidt. It follows that we can replace in the above Lemma $F$ by
$R(D_A)$ and we get
$$sign(D_A-\l) - h_A\epsilon h_A^{-1}\in L_2,$$
where $h_A$ is constructed as above with $x=R(D_A)_{21}.$ This condition
guarantees that $A\mapsto h_A$ is a section of $P$ and so we may define the
cocycle $\o(g;A)=h_{g\cdot A}^{-1}T(g)h_A$ as in section 3.

Near the unity in $\Cal G$ the extension $\GG$ behaves like the product $\Cal G
\times Map(\Cal A,S^1).$ It is convenient to consider the larger group with
the complexified fiber $Map(\Cal A,\CM^{\times}).$ The product is given then
by the formula
$$(g,f)(g',f')=(gg',f''), \text{ with } f''(A)=f(g'\cdot A)f'(A)\O(g,g';A),
\tag4.5$$
where the cocycle $\O$ is
$$\O(g,g';A)=\O_1(\o(g;g'\cdot A), \o(g';A)).\tag4.6$$

As a vector space, the Lie algebra of $\GG$ is the direct sum of $Map(M,\gm)$
and $Map(\Cal A,i\RM).$ The latter part is an abelian ideal, the infinitesimal
gauge transformations acting on it through Lie derivatives. Taking derivatives
of (4.6) along 1-parameter subgroups of $\Cal G$ one gets:

\proclaim{Theorem 4.7}The commutator
of two infinitesimal gauge transformations $X,Y$ is the pointwise commutator
plus the cocycle term $c_2(X,Y;A),$ computed through
$$\align c_2(X,Y;A)&=c_1(d\o(X;A),d\o(Y;A))\\&=c_1\left(h_A^{-1}T(X)h_A-h_A^
{-1}\Cal L_X h_A, h_A^{-1}T(Y)h_A-h_A^{-1}\Cal L_Y h_A\right).\endalign$$

\bf Remark. \rm In the $1+1$ dimensional case one may take $h_A\equiv 1$ and
then $c_2(X,Y;A)=c_1(T(X),T(Y))$ does not depend on the potential. When $M=
S^1$ this expression reduces to the usual formula for the central term of a
Kac-Moody algebra, [PS].

\bf Generalizations. \rm The method of this paper can be used in principle also
for 1) the group of local spin transformations , 2) for diffeomorphisms.
In the former case the base space of the Fock bundle is the space of spin
connections and in the latter the space of metrics of the physical 3-space.
In both cases the base space is flat and so the global trivializations
corresponding to the function $h_A$ above exist. The cocycles are then
constructed
as in the case of gauge transformations. However, I have not been able to
compute explicite formulas for the cocycles.

\bf Remark \rm There is an ambiguity in the cocycle $\O$ which is due to the
freedom to choose the local section for the bundle $\GG\to\Cal G.$ A change
in the local section is effected by a function $g\mapsto \b(g;A)$ from $\Cal G$
to the group $Map(\Cal A, S^1).$ The (local) 2-cocycle is then transformed by
a coboundary to the new cocycle
$$\O'(g,g';A)= \b(gg';A)^{-1}\b(g;g'\cdot A)\b(g';A) \O(g,g';A).\tag4.8$$
Thus only the cohomology class of $\O$ has invariant meaning. The cocycle
constructed above is nonlocal. There is another cocycle which is local; its
infinitesimal form was given in [M1],[F]:
$$(d\O)(X,Y;A)= \frac{1}{24\pi^2} \int_M \TR\, dA(XdY-YdX),\tag4.9$$
where the trace is evaluated in the representation $\rho$ of $G.$ This form
of the commutator anomaly can be derived from the Atiyah-Singer index theory
approach to
determinants of Dirac operators, [AS], as explained in [M3]. It appears also
in perturbation series computations in a Yang-Mills-Dirac system, [JJ].
A modern scattering theoretic treatment (in time dependent external fields)
can be found in [IO].  It would be interesting to have a "simple"
nonperturbative regularization $h_A$ which would directly lead to the nice
local form (4.9).

An interesting open problem is what form would the Sugawara construction of
a Hamiltonian take in  $3+1$ dimensions.  An early attemp to put together
operator valued Schwinger terms and  Sugawara's idea was made in [DG].

\vskip 0.5in
REFERENCES

\vskip 0.3in
[A] Araki, H: Bogoliubov automorphisms and Fock representations of
canonical anticommutation relations. In: \it Contemporary Mathematics, \rm
American Mathematical Society vol. 62 (1987).

[AS] Atiyah, M. and I. Singer: Dirac operators coupled to vector potentials.
Proc. Natl. Acad. Sci. USA \bf 81, \rm 2597 (1984). I. Singer: Families of
Dirac operators with applications to physics. Asterisque \bf 323 \rm (1985).

[DG] Dicke, A. and G. Goldin: Writing Hamiltonians in terms of local currents.
Phys. Rev. \bf D 5, \rm 845 (1972).

[F] Faddeev, L. : Operator anomaly for the Gauss law. Phys. Lett. \bf 145 B,
\rm 81 (1984). Also: L. Faddeev and S.L. Shatasvili: Algebraic and hamiltonian
methods in the theory of non-abelian anomalies. Theor. Math. Phys.
\bf 60, \rm 770 (1984).

[IO] Itoh, T. and K. Odaka: A particle-picture approach to anomalies in
chiral gauge theories. Fortschritte der Physik \bf 39, \rm 557 (1991).

[JJ] Jackiw, R. and K. Johnson: Anomalies of the axial vector current.
Phys. Rev. \bf 182, \rm 1459 (1969).

[K] Kac, V.: \it Infinite Dimensional Lie Algebras. \rm Cambridge University
Press, Cambridge, UK (1985).

[L] Langmann, E.:  Fermion and boson current algebras in (3+1)-dimensions.
In: \it "Topological and Geometrical Methods in Field Theory", \rm
eds. J. Mickelsson and O. Pekonen. World Scientific, Singapore (1992).

[Lu] Lundberg, L.-E.: Quasi-free "second quantization". Commun. Math. Phys.
\bf 50, \rm 103 (1976).

[M1] Mickelsson, J.: Chiral anomalies in even and odd dimensions.
Commun. Math. Phys. \bf 97, \rm 361 (1985).

[M2] ----------:   Kac-Moody groups, topology of the Dirac determinant bundle,
and fermionization. Commun. Math. Phys. \bf 110, \rm 173 (1988).

[M3] ----------: \it Current Algebras and Groups. \rm Plenum Press,
New York and London (1989).

[M4]  --------: Commutator anomaly and the Fock bundle. Commun. Math. Phys.
\bf 127, \rm (1990); On the Hamiltonian approach to commutator anomalies in
$3+1$ dimensions. Phys. Lett. B \bf 241, \rm 70 (1990).

[M5] ---------: Bose-Fermi correspondence, Schwinger terms, and Fock bundles
in $3+1$ dimensions. In: \it "Topological and Geometrical Methods in Field
Theory", \rm
eds. J. Mickelsson and O. Pekonen. World Scientific, Singapore (1992).

[MR] Mickelsson, J. and S. Rajeev:  Current algebras in $d+1$ dimensions and
determinant bundles over infinite-dimensional Grassmannians. Commun. Math
Phys. \bf 116, \rm 365 (1988).

[NA] Nelson, P. and L. Alvarez-Gaum\'e: Hamiltonian interpretation of
anomalies. Commun. Math. Phys. \bf 99, \rm 103 (1985).

[Pi] Pickrell, D.: On the Mickelsson-Faddeev extensions and unitary
representations. Commun. Math. Phys. \bf 123, \rm 617 (1989).

[PS] Pressley, A. and G. Segal: \it Loop Groups. \rm Calarendon Press, Oxford
(1986).

[R] Ruijsenaars, S.N.M.: Index formulas for generalized Wiener-Hopf operators
and Boson-Fermion correspondence in $2N$ dimensions. Commun. Math. Phys.
\bf 124, \rm 553 (1989).

[S] Segal, G.: Faddeev's anomaly in the Gauss' law. Preprint (unpublished),
Oxford (1985).
\enddocument